# Emerging AI Approaches for Cancer Spatial Omics

Javad Noorbakhsh[1], Ali Foroughi pour[2], Jeffrey Chuang[1,3]

[1] The Jackson Laboratory for Genomic Medicine, Farmington, CT

[1] St Jude Children's Hospital, Memphis, TN

[1] UCONN Health, Department of Genetics and Genome Sciences, Farmington, CT

## Abstract

Technological breakthroughs in spatial omics and artificial intelligence (AI) have the potential to transform the understanding of cancer cells and the tumor microenvironment. Here we review the role of AI in spatial omics, discussing the current state-of-the-art and further needs to decipher cancer biology from large-scale spatial tissue data. An overarching challenge is the development of interpretable spatial AI models, an activity which demands not only improved data integration, but also new conceptual frameworks. We discuss emerging paradigms – in particular data-driven spatial AI, constraint-based spatial AI, and mechanistic spatial modeling -- as well as the importance of integrating AI with hypothesis-driven strategies and model systems to realize the value of cancer spatial information.

## Background

Recent advances in highly multiplex spatially resolved omics (SRO), such as spatial transcriptomics and proteomics, have led to an explosion of studies on tissue spatial structure and its cellular underpinnings. Such approaches have the potential to revolutionize the histopathologic and molecular understanding of cancer. However, spatial data demand new analysis methods and concepts to address growing challenges in interpretability and reproducibility. Such challenges arise from the vastness of the data space spanning the intricate, yet largely undefined, spatial phenotypes within diverse tissue samples. Novel paradigms for biological discovery are needed to realize the translational value of these rich spatial resources.

SRO data are growing rapidly through imaging technologies, including stain-based (e.g. H&E, IHC), molecular mass-spectrometry (e.g. MALDI), transcriptomic (e.g. Visium, VisiumHD, Xenium, CosMx), and antibody-based proteomic (e.g. CODEX, CellDive) methods. Low-plex stain-based methods have been widely used in clinical settings for decades. Mass-spectrometry methods are versatile for measuring diverse molecular species, though with associated challenges in specificity. Most prominently, spatial transcriptomic and proteomic methods have accelerated in the past few years, providing high-plex, highly specific quantifications of RNAs and proteins at resolutions of ~0.25-100 microns. These spatial approaches extend single cell and bulk gene-profiling technologies that have been used widely in the last decade.

Due to the diversity and complexity of spatial phenomena within tissues, data-driven artificial intelligence approaches, in addition to mechanistic models, will be valuable to obtain insights from SRO data. Large-scale datasets are critical for AI-based data mining, and the scientific community will rely on spatial data being organized in computationally efficient, reusable, and standardized ways.  Analogous organizational pressures arose during the high-throughput sequencing revolution, with cancer consortium projects (e.g. TCGA, PCAWG) driving demand for standardization in datatypes (FASTQ, binary alignment: BAM) and genomic annotations (genes, variants, expression), centered around the goal of discovering driver mutations or expression states.  More recent consortia (Human Cell Atlas, HuBMAP [1] HTAN [2], SenNet [3], et al) and data aggregation initiatives (e.g CROST [4] and STOmicsDB [5] ) have been important for spatial omics data. However, the core goals of tissue spatial analysis have not yet been agreed upon. Despite the community's extensive experience with bulk and dissociated sequencing and protein data, effective goal setting will require openness to new conceptual paradigms.

The most prevalent paradigm for spatial omics analysis has been to extend approaches from single cell analysis, i.e. first, to aggregate sequence (or protein) data for each cell, and second, to analyze how expression relates to cell location; however, this paradigm has inherent limitations. As an illustration, pathologists often make clinically valuable decisions from histology images, which contain no sequence data. Some cancer pathology evaluations may not even depend on cells, for example relying on the morphology and density of blood vessels, the extracellular matrix, or necrotic cavities. Thus, cell-based approaches contain only part of the information valuable in tissue images. To determine the tissue features most relevant to patient outcomes, it will be vital to study not only the value of different data types, but also how to best encode SRO data into useful data representations. The choice of data representation (e.g. cell-based, graph-based, tile-based) limits the biophysical processes that can be studied, but such limitations have so far been little investigated.

In the following sections we discuss key conceptual challenges and possibilities for the understanding of spatial data, including the role that AI can play.  While such topics pertain broadly to tissue biology, spatial profiling has special translational value for cancer. Spatial relationships are critical to treatment response, for example by mediating the interactions of immune and cancer cells during immunotherapy. We conclude by discussing the need for tissue studies within perturbable model systems, which are needed to verify mechanistic understanding and pave the way for clinical translation.

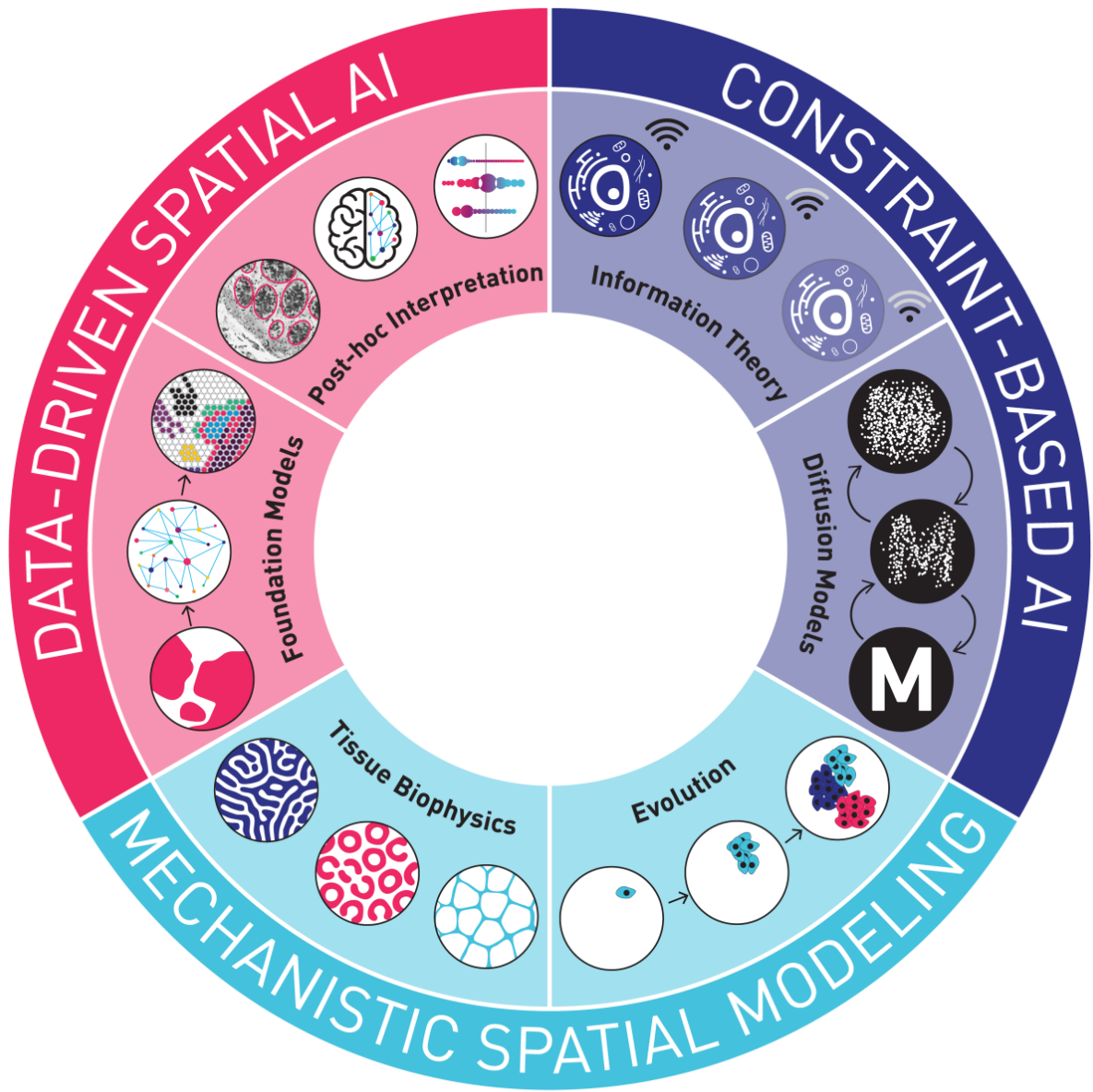

*Figure 1. Emerging analysis paradigms for cancer spatial AI.* **DATA-DRIVEN SPATIAL AI** *is a paradigm that avoids strong assumptions about the underlying data. For example,* **Foundation Models** *can be trained on large-scale histopathology or spatial omic data to complete diverse tasks at the level of histopathology, cellular networks, or individual cells. Foundation models typically involve deep neural networks with abstract feature spaces, so* **Post-hoc Interpretation** *is needed to identify the structures in the tissue predictive of cancer outcomes. A second paradigm is* **CONSTRAINT-BASED SPATIAL AI***, in which intuitive concepts are used to constrain model inference.* **Information Theory** *is an important framework to guide such models, as cellular interactions in cancer are constrained by information transfer.* **Diffusion Models** *are constraint-based AI models used widely in generative image AI and increasingly in tissue analysis, based on the presence of spatial hierarchies in images.* **MECHANISTIC SPATIAL MODELING** *is a paradigm based on hypothesis-testing and mechanistic discovery.* **Tissue Biophysics** *approaches integrate spatial biophysical concepts (e.g. reaction diffusion, tissue mechanics) with empirical spatial omics observations.* **Evolution** *is fundamental to the development and treatment response of tumors, and new spatial omics-based approaches are improving the quantification and understanding of this process.*

# Main text

## Emerging analysis paradigms for cancer spatial AI

The field of spatial omics is in the early stages of establishing a hierarchy of goals, a prerequisite for the evaluation of data collection and analysis approaches. Nevertheless, several concepts have begun to emerge. Here we review approaches of growing importance, which we organize into three paradigms: Data-driven Spatial AI, Constraint-based Spatial AI, and Mechanistic Spatial Modeling (Figure 1).

## Data-driven spatial AI

Many recent spatial analysis algorithms have been data-driven, i.e. they seek to identify patterns in data without pre-specifying biological hypotheses. Such approaches have the advantage of flexibility but can be difficult to interpret mechanistically, particularly when they use deep neural networks to embed biological data into abstract latent spaces. We consider three topics among the data-driven approaches: histopathology foundation models, spatial omic foundation models, and post-hoc interpretation.

### *Histopathology foundation models*

Histopathology foundation models have become a leading category of spatial machine learning model, typically involving deep neural network transformer architectures trained on whole slide hematoxylin and eosin (H&E) stained images (Figure 1, Foundation Models). Foundation models are large machine learning models trained on vast datasets, enabling them to tackle diverse tasks. For example, a single foundation model may enable image classification, segmentation, and annotation. In digital and computational pathology, these models have been trained on large H&E slide datasets, in some cases more than one million whole slide images, and have demonstrated value for integration into diagnostic workflows. Such foundation models encode images into data representations which can be used for disease classification, cell segmentation, and outcome prediction. Other clinically-driven image modalities, such as immunohistochemistry (IHC), have also been studied using foundation models [6], but H&E models are the most well-developed, due to the prevalent use of H&E by pathologists for clinical decisions.

Histopathology foundation models are based on deep neural network architectures applied to tiles within H&E whole slide images. Slide representations are then computed by aggregating the behaviors of the tiles. Early foundation models were based on deep convolutional networks, but most recent models utilize self-supervised training with a vision transformer [7] backbone. One popular implementation is DINOv2 [8], which seeks to distinguish visual features without model fine-tuning. This architecture has been used in recent foundation models including Virchow [9], UNI [10], and GPFM [11]. Histopathology foundation models have become progressively easier to use, for example having been integrated into end-to-end workflows such as STAMP [12], which facilitates the preprocessing of whole slide H&E images, model training, and evaluation in a single framework.

The wide success of large language models (LLMs), such as GPT [13], BERT [14], and LLaMA [15], has also sparked approaches to merge these with imaging tasks. A recent category of foundation models uses paired H&E images and text (e.g. CONCH [16], PathChat [17], and TITAN [18]). These approaches combine vision transformers and LLMs to generate integrated data representations and enable chat-based interrogation of histology images.

*Spatial omic foundation models*

Spatial omics data analysis has benefited from neural network approaches for many tasks, e.g. cell type annotation, batch correction, resolution enhancement, clustering, spatially variable gene detection, dropout imputation, and ligand-receptor detection [19], but general spatial omic foundation models are only beginning to be developed (Figure 1, Foundation Models). It is instructive to compare to dissociated single-cell-based foundation models, which are simpler yet still growing rapidly. For example, a wave of tools leveraging transformers has emerged for single cell deep learning [20]. A key innovation has been in the tokenization step, which specifies what aspects of the input data are important, e.g. gene identity, expression value, ranking, and metadata [21]. For instance, in scGPT [22] tokens comprise genes, expression quantile orders, and experimental conditions. scBERT [23] uses gene2vec [24] to embed co-expression patterns into gene tokens, while binning the expression values. scGPT, scFoundation [25], and CellPLM [26] are foundation models whose resultant embeddings, whether fine-tuned or direct, can be used as inputs for various tasks. Other models include tGPT [27], xTrimoGene [28], TOSICA [29], and Geneformer [30], and others [20]. A caveat of these foundation models is that the available pretraining datasets have been much smaller than for H&Es and LLMs [20], so caution in accepting their outputs is warranted. Such models should improve as the field matures through improved data, curation, and architectures.

Techniques devised for scRNA foundation models are beginning to be extended to spatial models. Such models also typically rely on transformer architectures trained on large datasets using self-supervision. A few of these models have been introduced as foundation models and can enable multiple downstream tasks. In particular, CellPLM [26] is a transformer encoder/decoder VAE which can do cell clustering, denoising, imputation, and cell type annotation, while Nicheformer [31] is a transformer model capable of label transfer and prediction of cellular neighborhood compositions. Both methods use a combination of spatial and dissociated single cell RNA sequencing (RNAseq) as input, which potentially boosts their ability to transfer information across the two data modalities. Other models such as SpaFormer [32], and stEnTrans [33] are transformer models trained on spatial transcriptomics data which could potentially be utilized as foundational.

Foundation models for spatial proteomics are also a growing area. Spatial proteomic technologies have facilitated in-depth investigation of the tumor microenvironment (TME), offering insights into tumor dynamics and its interactions with the immune system [34] [35].

Advances in multiplex techniques, such as CODEX [36] [37], IMC [38], and Cell Dive [39], have allowed simultaneous measurement of many proteins at subcellular resolution. Such data can enable direct modeling of biophysical processes, including fine interactions such as synapse formation between cells [40]. However, integrating protein data across experiments remains challenging due to variations in protein panels and measured antibody intensity distributions. These issues may be alleviated by encoding protein data into more robust representations, e.g. by non-negative matrix factorization [41] or generative neural networks, which have also enabled powerful capabilities such as combinatorial protein signal decomposition [42] [43]. A promising recent method named KRONOS [44] addresses the issue of protein marker heterogeneity by an innovative tokenization approach. The authors pass the marker identity information as a secondary positional encoding and effectively treat all markers as equivalent otherwise. Their model outperformed existing foundation models that are trained on out-of-domain histology data or multichannel cell profiling images. Still, such encodings may create non-intuitive statistical artifacts, and broader research into proteomic foundation model development and their benchmarkings are important to the further development of the field.

*Post-hoc interpretation*

Despite their impressive capabilities, foundation models are often criticized as 'black box' systems because their causal logic is not directly interpretable [45] [46]. The loss of interpretability arises from the embedding of image data into abstract latent spaces using deep neural networks. Interpretation of the genetic and cellular features associated with foundation model image embeddings therefore must be *post hoc,* i.e. based on data patterns rather than causal relationships among biological entities (Figure 1, Post-hoc Interpretation). This is in contrast with mechanistic models, which are formulated from experimentally controllable features such as genes or cells.

Some architectures, such as transformer models, provide their *post hoc* interpretability through the attention mechanism, which highlights what input features are focused on by individual attention-heads within the neural network. However, inter-head correlations are difficult to interpret, and attention weights do not uniquely translate to output feature importances [47] [48] [49]. Explainable AI (XAI) techniques like SHAP [50] and LIME [51] address some of these limitations by attributing model predictions to individual features. SHAP uses a game-theoretic strategy treating features as cooperative players contributing to the prediction, while LIME builds a simpler explainable model through local perturbations of the original model around a data point of interest. In addition to model perturbation, data perturbation can be used to identify biologically plausible perturbations through context-guided data generation [52]. Additionally, methods inspired by physical entropy and thermodynamics have been proposed to identify optimal explanations for model embeddings [53]. These techniques can help capture spatial structures in images that are most indicative of outcomes.

Other efforts have been made to develop interpretable models of features related to tissue organization. A useful concept is the functional tissue unit (FTU) [54], defined as the smallest multicellular tissue unit performing a specific function within its microenvironment that is replicated in a whole organ [55]. Repetitive FTUs can be captured by spatial frequency analysis of SRO data. For example, SpaGFT converts the SRO data into a graph and applies a Graph Fourier Transform to identify such features. This method can be implemented as an explainable regularizer for other machine learning models, improving their interpretability [56]. Such modular approaches based on FTUs, even if defined post hoc, are likely to be increasingly used in SRO analysis to demonstrate robustness and reusability of computational frameworks.

Current spatial proteomics data have lower plexity than spatial transcriptomics data, so visual inspections of individual proteins, cell types, and cellular neighborhoods are often used to interpret protein contributions to cancer survival risk [41] [57] [58] . However, as technologies for proteomic plexity and throughput improve, deep learning and interpretability considerations will grow for these data types as well [43].

## Constraint-based spatial AI

Alternative paradigms based on constraints on the SRO data representation can provide more interpretability than foundation models.  For example, all tissue structures are constrained by biophysical processes, but the underlying cellular mechanisms are variable and complex.  AI approaches that account for these constraints abstractly, without focusing on individual mechanisms, have grown. We discuss such approaches for cancer, focusing on those incorporating information theoretic constraints and those that mimic spatial constraints, notably image diffusion models.

### *Information theory-based constraints*

Information theory is a mathematical framework for quantifying the flow, processing, and storage of information within systems, and it can be applied to guide spatial analysis in tumors.  For example, information theoretic concepts can be used to quantify limits on tissue heterogeneity, signaling dynamics, and other types of spatial organization, such as intratumoral immune infiltration and epithelial-to-mesenchymal transitions. Such approaches are helpful for understanding the bounds on inference for different models (Figure 1, Information Theory).

Spatial information transfer in tissues has fundamental limits. This has been demonstrated for drosophila embryos, where morphogen gradients have been optimized by evolution to the physical limits of signal transduction [59]. Such evolutionary optimization likely extends to complex tissues, where information transfer is governed by physical and biological constraints that have been shaped by developmental processes. Viewing this as an information theory problem, with channel capacity and mutual information defining these constraints, provides a powerful analytical framework. In cancer, deviations from these optimal solutions may result in identifiable patterns in SRO data, revealing novel mechanisms driving cytokine, endocrine, and immune signaling [60] [61]. Quantifying signal propagation and interaction provides a way to understand collective cellular behaviors within cancer tissues driven by external or internal cues.

Spatial constraints on information transfer can motivate SRO dimensionality reduction approaches. For example, non-negative matrix factorization (NMF) has been used to reduce glioma expression data into interpretable patterns of spatially co-expressed genes, i.e. "metaprograms," [62], though it remains challenging to know what spatial scales are appropriate for such approaches. Alternatively, the information bottleneck (IB) method [63] offers an information-theoretic approach to attain the most compressed, lower-dimensional representation of a high-dimensional dataset that is maximally predictive of a desired outcome. bioIB [64] has developed this concept for scRNAseq data to determine a set of genes ('meta-genes') which are predictive of disease status or cell type. Extending such approaches to the spatial domain offers new opportunities for developing more robust SRO data analysis tools.

Extensions of IB have been applied to deep learning models, where the layers in the model are treated as an information channel, and IB is applied to each layer [65] [66]. These layers progressively represent a desired outcome, with each layer 'forgetting' certain input details to better learn the output. Such methods could be extended to spatial data through architectural designs such as graph neural networks (GNN), convolutional neural networks (CNN), or Vision Transformers (ViT). This type of strategy may elucidate how tissue-level information flows through molecular and spatial interactions, enabling the extraction of minimal yet predictive feature sets at multiple spatial scales. These minimal encodings could identify molecular signals and spatial arrangements for intercellular communication, revealing key length scales and structural patterns that govern tissue organization and behavior.

*Image diffusion models*

Diffusion models guided by theoretical frameworks that incorporate tissue-specific symmetries (e.g., repetitive structures, rotational and translational invariance) are a promising approach for spatial analysis (Figure 1, Diffusion Models). These models have achieved significant advances in generative image AI, including widely used tools like DALL-E [67] and Stable Diffusion [68]. Diffusion models are comprised of an encoder that iteratively degrades an image's data distribution structure through a diffusion-like process, followed by a decoder that reverses the process to reconstruct the image with progressive detail. Training such an encoder-decoder on image datasets yields a neural network able to generate images from noise [69]. Latent diffusion models [70] extend traditional diffusion approaches by first reducing the dimensionality of input data before applying the diffusion process to the compressed representation. This strategy enhances computational efficiency and has been integrated with transformers, forming diffusion transformers (DiTs) [71].

Diffusion models are now being applied to spatial transcriptomics. For example, stDiff [72] and SpaDiT [73] utilize DiTs to impute missing genes in scRNAseq data and spatial transcriptomics. However, a key limitation of these models is their lack of spatial coordinate integration during training, which restricts their ability to fully capture and utilize spatial relationships. Some diffusion models integrate spatial information to better represent the unique characteristics of spatially resolved data. For example, DiffuST [74] employs latent diffusion models alongside graph autoencoders to resolve semantic inconsistencies across data modalities while capturing spatial relationships. SpatialDiffusion [75] incorporates spatial coordinates, gene expression, and cell type information to predict unseen slices in 3D spatial transcriptomics, interpolating distributions from neighboring slices. Similarly, stMCDI [76] utilizes a graph neural network (GNN) to encode spatial information, leveraging a diffusion model to impute missing data in spatial transcriptomics.

The success of diffusion models is rooted in statistical thermodynamics, as they leverage concepts such as phase transitions, symmetry breaking, and critical instabilities to achieve accurate image reconstruction [77], offering a theoretically-supported approach for generating interpretable representations. While such models have been studied for images with few data channels, extending these principles to high-dimensional spatial omics presents a key challenge important for improving SRO data interpretation [78]. Furthermore, the iterative encoding process in diffusion models parallels the spatial hierarchies within images [79]. Investigation of this relationship may be valuable for improving interpretability. Another promising approach is context-constrained diffusion models [52], which can improve generation of biologically realistic data to facilitate interpretation of foundation models as well as provide augmented data during training.

## Mechanistic spatial modeling

Tissue spatial profiling data are, essentially, measurements of 3D materials. In materials science, macroscopic behaviors are studied as mechanistically deriving from microscopic processes such as electromagnetic and molecular interactions. Likewise, tumors are impacted by microscopic biophysical processes such as chemical signaling and diffusion, resulting in contiguous region types (tumor, immune-infiltrated, necrotic, fibrotic, etc.) with internally coherent cell composition, cellular states, or extracellular structures.  Some analysis approaches, e.g. GASTON [80], leverage this coherence to identify region types. However, most current approaches are empirical rather than grounded in biophysical processes. Integration of physics-inspired concepts is therefore a promising direction, and there is a critical need for new AI models capable of mechanistic inference from SRO data. Below we discuss approaches using spatial data to learn biophysical and evolutionary processes within tumors.

### *Inference of tissue biophysics*
Cellular dynamics within tissues can be viewed as a reaction-diffusion process (Figure 1, Tissue Biophysics), where cell-intrinsic mechanisms and cell-to-cell communication occur simultaneously with diffusion. Computational models of cancer based on this perspective have been developed [81] [82] [83], but have been limited in the number of cell types covered and have relied on partial differential equations with many unknown parameters. However, for a hypothesized model of tissue architecture, the governing equations should be learnable if sufficient data are available. A recent category of inference method for learning equations from large datasets is physics-informed neural networks (PINNs) [84], which embed physical equations into neural networks. These methods require dynamical data for training, and they have been applied to scRNA data to predict cell state dynamics [85] and for RNA velocity inference [86]. So far, application of PINNs to tissues has been limited due to scarcity of time-course data. However, as longitudinal SRO data improve, PINNs can provide a framework to infer interpretable physical processes and parameters from them. For example, HoloNet [87] is a graph neural network that uses ligand diffusion equations to infer ligand-receptor interactions from spatial transcriptomics data, and SpaCCC [88] integrates this into a transformer framework. These two methods overcome the limitations of time-course scarcity by stripping chemical diffusion and reaction equations from their dynamics and treating them as steady-state functional forms.

Drawing inspiration from these PINN approaches, it may be possible to model cancer tissue as governed by multiple fields mediated by the spatial diffusion of molecules. An SRO PINN model could be built around cells acting as relays of such fields—each with its own transmission length scale, which could be inferred from gene or protein expression. These fields could describe, for example, molecular densities or mechanical forces, and cross-attention between them could reveal their interactions, such as ligand-receptor activation. Knowledge of chemical diffusion may be directly incorporated into PINN architecture. For example, multilayer perceptron modules could be used to link gene expression to signal concentration or diffusion rate, while convolution-like kernels could approximate the diffusion process across different length scales. This approach could provide a framework to quantify the spatial dynamics of immune, stromal, and cancer cell populations, as well as their modulation by host factors.

While most neural networks rely on multilayer perceptrons (MLP), which are feed-forward networks with learnable linear edges and fixed nonlinear nodes, Kolmogorov-Arnold Networks (KANs) [89] take the opposite approach, learning nonlinear functions on edges while keeping nodes linear. This design allows KANs to directly infer sub-functional components of a global function, making them potentially more interpretable. Also, although similar to PINNs in their ability to infer system dynamics, KANs are not limited to dynamic data. KANs integrated with convolutional neural networks have been successfully used in remote sensing applications [90] involving geographic spatial profiles at hundreds of light frequencies. Such data are analogous to spatial omics data in their high number of channels and environment-influenced spatial relationships. Thus, KANs may be effective for interpreting how groups of biomarkers and cell types form structural phenotypes within tissues, using the mathematical functions inferred along KAN edges.

Physical processes such as diffusion can also be qualitatively incorporated into neural networks. For instance, sepal [91] is a spatial transcriptomics analysis tool which assumes that transcripts diffuse in the environment according to a fixed diffusion rate. It then ranks the genes by the time required to reach homogeneity. Although in this method the diffusion process is not physically observed, this approach is nevertheless able to determine spatial structures and their related gene families.

Tissue mechanical properties such as stiffness, adhesion, and viscosity are also important to cancer. These impact tumor invasiveness [92] [93] [94], drug diffusion, and angiogenesis, likely mediated by the impact of mechanical properties on cellular polarity, membrane rigidity, and cell migration [95] [96]. SRO measurements have the potential to reveal mechanical parameters. For example, spatial transcriptomics data has been used to infer tissue stiffness by solving the equations of surface tension from segmented cell membranes [97]. Although the use of cancer SRO data for tissue mechanics studies has been uncommon, the substantial literature on cancer tissue biophysics [98] [99] suggests that such approaches have untapped potential [100].

*Evolution*

Evolution has been extensively used to interpret tumor formation, heterogeneity, treatment response, and resistance (Figure 1, Evolution). Many studies have incorporated scRNAseq data toward understanding cancer evolution [101], and analogous spatial transcriptomics-based analysis are growing [102]. Because evolution is classically defined by genotypes, a common use of SRO data has been to optimize spatial genotype calling. Methods like InferCNV [103], which was developed to infer copy number variations (CNV) from scRNAseq data, are now regularly applied to spatial transcriptomics data [104] to identify local CNV-defined genotypes. Analogously, transformer-based CNV-calling methods like CoT [105], which has been used for genome-wide denoising of single cell DNA sequencing (scDNAseq), could be adapted for spatial DNA sequencing. Integrating spatial continuity into these models can improve subclonal phylogeography inference. CalicoST [106] does this, including inference of allele-specific copy numbers, and it has been used to identify subclonal heterogeneity as well as oncogenic and metabolic activity in HTAN datasets [102]. Currently, CalicoST does not leverage long-range spatial correlations, focusing on spot neighborhoods. However, long-range correlations could arise in images from geographic selection pressures within the tumor, or due to the limited ability of 2D images to represent 3D spatial processes. Improved attention-based networks will be valuable to better capture these effects. PINNs have also been applied to capture tumor growth dynamics [107]. Combining these methods with genomic-based methods may improve phylogeography inference from SRO data.

A broader consideration for tumor evolution is that epigenetic and morphological changes occur together with genotype evolution. For example, epigenetic shifts in tumor cells are important to treatment resistance [104]. Therefore, simultaneous integration of spatial gene expression and phylogenetics in a single network would be valuable. This could be achieved by feeding expression and genotype matrices into a unified tokenization scheme. Tissue morphology, as characterized by H&E and analyzed via foundation models, has also been shown to predict some tumor expression states and genotypes [108] [109]. This suggests that it may be possible to improve inference of tumor evolutionary processes by embedding local genomic data together with the tissue morphology and transcriptional gradient context.

## Data considerations

### Data integration

Cancer multimodal spatial data are diverse [102], creating data integration opportunities and challenges. At one end, H&E data are widely available but low plex. On the other end, spatial omics datasets are scarce but can have thousands of markers. It remains poorly understood what information is common across modalities and what is modality-specific. To benefit from spatial data integration, technical considerations must be addressed. Different modalities require distinct scaling, standardization, and filtering to avoid biasing downstream models. Image registration across modalities is challenging due to differences in data formats, mismatched markers, resolutions, sample preparation, tissue size, and batch effects [110]. Measurement noise, including technical and biological variability, can complicate data interpretation [111]. Spatial RNA can be affected by platform-specific dropout and probeset limitations [112].

Nevertheless, spatial data integration is already benefiting cancer research. For example, cancer H&E images have been shown to predict spatial expression-identified patterns including immune infiltration and drug-induced persistence [104] [113] [114]. Pipelines are available to ease H&E/spatial transcriptomic integration for cancer, such as STQ [115]. Other general integration methods will also be useful for cancer. For example, SpatialGlue [116] integrates spatial transcriptomics, proteomics and epigenomics data for domain identification. Architectures based on single-cell integration frameworks such as scGPT [22] and scMoFormer [117] provide a model for expanding spatial data integration. COVET [118] uses localized expression covariances to encode cellular niches in spatial transcriptomics data. ENVI [118] is a variational autoencoder that integrates with COVET to simultaneously infer missing genes in spatial transcriptomics data inferred from dissociated single cell data, while assigning niche information to dissociated cells using spatial data. Approaches based on large-scale H&E integration with expression data are also of growing value [119]. For example, OmiCLIP [120] encodes highly expressed genes from local spots into sentences, then uses vision-language deep learning theory to build an H&E-omics foundation model.

### Mechanism-driven data generation

Spatial data have tremendous potential for improving prediction of clinical outcomes (e.g. [121]) while simultaneously suggesting new mechanisms. However, post-hoc interpretation approaches conflict with hypothesis-oriented standards for mechanistic discovery. For example, cancer clinical H&E samples have been studied extensively by foundation model approaches, but it remains unclear how to ascertain causal relationships from them. Patient IRB considerations also limit the types of clinical samples that can be obtained. Such issues will constrain human SRO-based foundation model approaches, even if large cancer datasets can be amassed [122]. To verify causal mechanisms, it will be important to purposefully generate spatial data for mechanistic discovery.

Mice are the canonical mammalian model organism, making them ideal for the study of cancer spatial mechanisms in a tissue context. The ease of mouse genetic engineering and organismal perturbation has a long history of enabling hypothesis-based discoveries of oncogenes, tumor suppressors, metastatic processes, tumor environmental interactions, and effects of aging [123]. Sophisticated mouse population genetic systems such as the Diversity Outbred and Collaborative Cross [124] have also revealed genes important to cancer mechanisms [125] and quantitative trait loci predictive of cell morphology [126]. Dynamic SRO data can be generated more easily in mice than from clinical samples, a key need for the training of PINNs. Stated simply, mice and other organismal models enable hypothesis-driven science on tissues, addressing the central challenge of post hoc mechanistic interpretation.

To realize the value of mouse and other animal model SRO cancer studies, such knowledge must be transferable to human. Identification of orthologous behaviors between mouse and human has not yet been well-quantified with SRO data, though projects such as the Cellular Senescence Network (SenNet) [3] are performing spatial profiling of tissues across some matched mouse and human organs. Methods for the scRNA version of this problem are being actively developed. For example, CAME [127] aligns scRNAseq data across species and enables the transfer of cell type labels, which it accomplishes via a graph neural network that integrates a gene-gene graph of homologous genes and a cell-cell graph of transcriptionally similar cells. BrainAlign [128] extends this idea to spatial transcriptomics by adding a graph linking spatially proximal spots on the tissue. BrainAlign can align human and mouse brain tissue as well as identify conserved and species-specific gene expression patterns. Existing SRO transformer models may also be developed for human and mouse SRO comparisons. For example, Nicheformer [31] uses a unified gene-based tokenizer for human and mouse spatial transcriptomics and single cell RNAseq, producing embeddings for tasks such as spatial label prediction. However, it does not use spatial coordinates as input. In contrast, SpaFormer [32] uses a cell-based tokenizer and explicitly encodes positional information to impute missing spatial transcriptomics data. A combined architecture with multi-species tokenizers and explicit positional encodings could better align human and mouse SRO data. All of these approaches will require further development of expert annotated SRO sets in mouse and human to train cross-species spatial aligners.

# Conclusions

The field of spatial omics is expanding rapidly in cancer research due to the critical importance of location-dependent interactions of cells within the tumor microenvironment. Cancer spatial omics datasets are high-dimensional and diverse, necessitating improved analytical paradigms. We have described three major paradigms for the development of the field: data-driven spatial AI; constraint-based spatial AI; and mechanistic spatial modeling.

We have also reviewed key additional considerations in data integration and mechanism-driven data generation.

While the scale and complexity of SRO data create strong demand for AI-based approaches, deep neural networks have only post-hoc interpretability compared to classical hypothesis-oriented approaches. Among the three major paradigms, deep neural network-based data-driven approaches, such as foundation models, are the least interpretable or mechanistic. Constraint-based spatial AI, such as image diffusion models, improve on this by incorporating biologically reasonable constraints into the underlying neural networks. Mechanistic models are the most interpretable, as they are constructed based on experimentally perturbable entities such as cells or genes. PINNs are a particularly promising approach for mechanistic modeling, as they combine explicit mathematical modeling of perturbable entities with data-driven inference amenable to SRO data. This provides a framework to jointly study biophysical and cell biological processes within tissues. Such joint investigation is essential to the underlying spatial processes of cell motility, signaling, replication, and evolution within the tumor microenvironment.

A general challenge for spatial omics foundation models is validation. Although tests such as tumor-type classification are commonly accepted for H&E foundation model benchmarking, spatial omics foundation model benchmarking is not yet standardized. To be useful, SRO foundation models should be able to distinguish fine microenvironments within tumors, a more complex task that is also inherently multiscale. Thus, SRO foundation model benchmarking will require community agreement on standardized evaluation tasks at multiple spatial scales. Alternative statistical models may be valuable in circumventing these challenges, e.g. constraint-based statistical models can be more interpretable and require less training data than deep learning counterparts [129]. Models based on message passing are a canonical example, where information of neighboring cells are aggregated to represent the cellular composition and spatial relations [130]. These models share similarities with GNNs, but require less training data. Interestingly, recent studies have combined GNNs with message passing techniques to generate improved embeddings given small data [131] [132] [133].

While spatial data are growing for clinical cancer samples, clinical data are restricted by regulatory and collection limitations. These restrictions make it difficult to generate the perturbative data important for hypothesis-based science or the timecourses important for training of PINNs. Mouse models can address this problem, as biophysical and genetic perturbations are possible with mice, and timecourse data are easier to generate with mice than in the clinic. Still, mice may respond differently to treatment than patients, and SRO-based models trained on mice will likely need to be fine-tuned on human data. Eventually, curated repositories that integrate human and mouse SRO cancer tissue data will be vital for the field. These will enable improved identification and validation of functional tissue units, reusable data analysis, and more precise delineation of the spatial processes essential to cancer marker identification, drug targeting, and clinical translation.

## List of abbreviations

**AI** – Artificial Intelligence
**BAM** – Binary Alignment Map
**BERT** – Bidirectional Encoder Representations from Transformers
**CNN** – Convolutional Neural Network
**CNV** – Copy Number Variation
**CODEX** – CO-Detection by indEXing
**CONCH** – Context-aware Chat-based Histopathology
**DNAseq** – DNA Sequencing
**DiT** – Diffusion Transformer
**FTU** – Functional Tissue Unit
**GNN** – Graph Neural Network
**GPT** – Generative Pre-trained Transformer
**H&E** – Hematoxylin and Eosin
**HTAN** – Human Tumor Atlas Network
**IB** – Information Bottleneck
**IHC** – Immunohistochemistry
**IMC** – Imaging Mass Cytometry
**KAN** – Kolmogorov-Arnold Network
**LIME** – Local Interpretable Model-agnostic Explanations
**LLM** – Large Language Model
**MALDI** – Matrix-Assisted Laser Desorption/Ionization
**MLP** – Multilayer Perceptron
**NMF** – Non-negative Matrix Factorization
**PCAWG** – Pan-Cancer Analysis of Whole Genomes
**PINN** – Physics-Informed Neural Network
**RNAseq** – RNA Sequencing
**SHAP** – SHapley Additive exPlanations
**SRO** – Spatially Resolved Omics
**SenNet** – The Cellular Senescence Network
**TCGA** – The Cancer Genome Atlas
**TME** – Tumor Microenvironment
**ViT** – Vision Transformer
**XAI** – Explainable Artificial Intelligence
**scDNAseq** – Single-cell DNA sequencing
**scRNAseq** – Single-cell RNA sequencing

## Declarations

**Data availability:** Not applicable

**Competing interests:** The authors declare no competing interests

**Funding:** The authors acknowledge support from The Jackson Laboratory Cancer Center's Cancer Advanced Technology (CATch) program, as well as NIH grants R01 CA230031, U54 AG075941, and P30 CA034196

**Author's contribution:** JN and JC planned and drafted the original version of the manuscript. AFP reviewed, provided new ideas, and drafted revisions.

**Acknowledgements:** The authors thank Karolina Palucka, Brian White, Frederick Varn, Francesca Menghi, and Kevin Anderson for valuable discussions.